\documentclass[twocolumn,showpacs,amsmath,amssymb]{revtex4}
\usepackage{graphicx}

\newcommand{\bg}{\mbox{\boldmath $\gamma$}}
\newcommand{\bt}{\mbox{\boldmath $\theta$}}

\newcommand{\by}{\mbox{\boldmath $y$}}
\begin{document}

\title{Geometry and Dynamics of Emergent Spacetime from Entanglement Spectrum}


\author{Hiroaki Matsueda}
\affiliation{
Sendai National College of Technology, Sendai 989-3128, Japan
}
\date{\today}

\begin{abstract}
We examine geometry and dynamics of classical spacetime derived from entanglement spectrum. The spacetime is a kind of canonical parameter space defined by the Fisher information metric. As a concrete example, we focus on the spectrum for free fermions in spatially one dimension. The spectrum has exponential family form like thermal probability distribution owing to mixed-state feature emerging from truncation of environmental degrees of freedom. In this case, the Fisher metric is given by the second derivative of the Hessian potential that can be identified with the entanglement entropy. We emphasize that the canonical parameters are nontrivial functions of partial system size by the truncation, filling fraction of fermions, and time. Then, the precise determination of this nontrivial mapping is necessary to derive the functional form of the Hessian potential that leads to correct entanglement entropy scaling. By this potential, we find that the emergent geometry becomes anti-de Sitter spacetime with imaginary time, and a radial axis as well as spacetime coordinates appears spontaneously. We also find that the information of the UV limit of the original free fermions lives in the boundary of the anti-de Sitter spacetime. These findings strongly suggest that the Hessian potential for free fermions has enough geometrical meaning associated with gauge-gravity correspondence. Furthermore, some deformation of the spectrum near the conformal fixed point is mapped onto spacetime dynamics. The fluctuation of the entanglement entropy embedded into the spacetime behaves like free scaler field, and the dynamics is described by the Einstein equation with a negative cosmological constant. Therefore, the Einstein equation can be regarded as the equation of original quantum state.
\end{abstract}

\maketitle

\section{Introduction}

Classical geometrical representation of quantum field theory is recognized to be a milestone to construct quantum gravity theory. A well-known example is the anti-de Sitter spacetime / conformal field theory (AdS/CFT) correspondence~\cite{Maldacena1,Maldacena2}. Now our interest is to understand this kind of problem with the help of information theoretical concepts. Two of key concepts are quantum entanglement and holography. For instance, the entanglement entropy in the CFT side can be holographically mapped onto a geometrical object like a minimal surface in the classical AdS side~\cite{Takayanagi1}. An important viewpoint inherent in this mapping is how to storage quantum data into classical spacetime that reflects symmetry of the original quantum system.

Recently, it has gradually been recognized that the information geometry may be also a powerful tool for examining the above problem~\cite{Amari}. In this geometry, the Fisher metric plays a central role on deeply understanding quantum/classical or micro/macro correspondence. This is because the Fisher metric creates a parameter space defined by a probability distribution (this space becomes spacetime when the distribution is time-dependent), and usually the distribution characterizes our starting quantum or microscopic model. In other words, this method naturally provides us with a correspondence from a microscopic system to classical geometry. Therefore, we can easily imagine that the method gives us information-theoretical interpretation of the AdS/CFT correspondence. Detailed examination of these arguments is the central purpose of this paper.

In more physical standpoints, geometrical approaches based on the Fisher information have not only been used for several optimization theories, but have been already applied to thermodynamics~\cite{Ruppeiner,Brody}. The probability distribution of thermodynamics is the Boltzmann distribution, which belongs to the so-called exponential family. In this case, the corresponding geometry becomes quite simple~\cite{Shima}, and we can see some physical meanings hidden in the geometrical representation. It has actually been found that the scaler curvature is related to the inverse of the free energy, when we regard inverse temperature and chemical potential as two independent coordinates in the parameter space. Since the free energy changes drastically near critical points, the curvature can detect criticality in the microscopic side. This feature reminds us of some similarity to the AdS/CFT correspondence in the sense that the classical side can capture criticality of the quantum side.

Furthermore, a close relationship between multiscale entanglement renormalization ansatz (MERA) and AdS/CFT correspondence has been also examined recently by the Bures or fidelity metric which is also basically equivalent to the Fisher information~\cite{Vidal1,Takayanagi4,Takayanagi5}. It has been found that the metric of the MERA network is actually AdS. The author has also examined the finite-temperature extention of the MERA network~\cite{Matsueda2} with great help of some notions of quantum entanglement, and this network structure coinsides with the black hole configuration in the AdS spacetime. These two findings are also important for understanding quantum/classical correspondence in terms of entanglement entropy and its holography.

A direct motivation of this work originates in entanglement thermodynamics in which we examine the first variation of the entanglement entropy~\cite{Blanco,Casini,Wong,Takayanagi2,Takayanagi3,Faulkner,Banerjee,Nima}. This idea leads to the so-called entropy-energy relation that is a kind of generalized first law of thermodynamics. It has been shown that the gravitational interpretation of the entropy-energy relation actually leads to the linearized Einstein equation. However, we will later see that the second derivative of the entanglement entropy, not the first derivative, is a crucial factor in the information geometry, and thus the previous results would be improved to find full Einstein equation~\cite{Matsueda1}.

Motivated by these previous works, we perform detailed analysis of geometric structure and dynamics of spacetime spanned by the Fisher metric with exponential family form of a probability distribution. In this case, the Fisher metric can be given by the second derivative of the Hessian potential that is basically equivalent with the entanglement entropy. Thus, finding a correct form of the Hessian potential in the exponential family representation is crucial.  As a typical example of CFT, we start with spatially one-dimensional (1D) free fermions, and we will see that their entanglement spectrum coinsides with the exponential family form~\cite{Vidal2,Peschel1,Peschel2,Peschel3,Cheong1,Cheong2,Lefevre,Jin}. At the same time, it is crucial to notice that the canonical parameters of the exponential family form are nontrivial functions of original model parameters (partial system size, filling fraction, and time). After this important finding, we can derive the correct form of the Hessian potential as a function of the canonical parameters, and can prove that the potential can be identified with the entanglement entropy. We then find that the Fisher metric becomes hyperbolic (AdS with imaginary time) after further general coordinate transformation. We emphasize crucial importance of finding correct coordinate transformations among original model parameters in the quantum side, canonical parameters in the exponential family form, and the final AdS metric. We will mention the presence of the concept of bulk/boundary correspondence in the present information geometry. Furthermore, when we consider some deformation of the spectrum near the conformal fixed point, the deformation is mapped onto dynamics of the spacetime, and this can be represented by the Einstein equation. We can introduce the fictitious energy-monentum tensor, and the tensor originates in the Lagrangian in which the free scaler field is equal to the embedded entropy data. Therefore, we argue that the Einstein equation in the emergent spacetime is a kind of equation of state of the original quantum system.

All of the present results strongly suggest that the information geometry is quite powerful to understand detailed structure of the AdS/CFT correspondence.

The organization of this paper is as follows. In the next section, we introduce the Fisher metric. In Sec.~III, we discuss about details of exponential family form and also mention that the entanglement spectrum for 1D free fermions consides with this form. In Sec.~IV, we examine geometric structure spanned by this metric, and find information-geometrical meaning of the Einstein equation in the emergent spacetime. The final section is devoted to the summary part.

\section{Fisher Metric}

In the theory of information geometry, we start with a probability distribution $\lambda_{n}(\bt)$ that depends on a discrete stochastic variable $n$ and an internal parameter set $\bt=\left(\theta^{1},\theta^{2},...,\theta^{D}\right)$. Later, we will discuss that the distribution is defined from entanglement properties of quantum systems. We are going to construct an information space spanned by the coordinates $\bt$, into which the data of $\lambda_{n}(\bt)$ are embedded. In convenience, we call this space as information space. However, in some cases, this can be a kind of spacetime, since the original data may be dynamical and time-dependent. Hereafter, we assume that we can differentiate $\lambda_{n}(\bt)$ by $\bt$ as many as possible. A well-known case is Gaussian in which the average and the variance are two internal parameters defining this distribution. The probability distribution obeys the following relation
\begin{eqnarray}
\sum_{n}\lambda_{n}(\bt)=1. \label{conservation}
\end{eqnarray}
For later convenience, we abbreviate the expectation value of a distribution function $O_{n}(\bt)$ by the angle bracket as
\begin{eqnarray}
\left<\mbox{\boldmath $O$}\right>=\sum_{n}\lambda_{n}(\bt)O_{n}(\bt), \label{avarage}
\end{eqnarray}
where we omit the index $\bt$ in the bracket and use the bold symbol. By defining the spectrum as
\begin{eqnarray}
\gamma_{n}(\bt)=-\ln\lambda_{n}(\bt), \label{gamma}
\end{eqnarray}
the information entroy is then given by
\begin{eqnarray}
S(\bt) = -\sum_{n}\lambda_{n}(\bt)\ln\lambda_{n}(\bt) = \left<\bg\right>. \label{entropy}
\end{eqnarray}
This is one of key parameters in this paper.

The Fisher metric is derived from the Kullback-Leibler divergence $D_{KL}$. The divergence measures difference between two similar probability distributions. The divergence is given by
\begin{eqnarray}
D_{KL}=\sum_{n}\lambda_{n}(\bt)\left(\gamma_{n}(\bt+d\bt)-\gamma_{n}(\bt)\right). \label{KL}
\end{eqnarray}
This divergence is not symmetric for the exchange between $\gamma_{n}(\bt)$ and $\gamma_{n}(\bt+d\bt)$, but the second-order of its Tayler expansion has better properties as a measure of the distance. Actually, we can derive the following result
\begin{eqnarray}
D_{KL} = \frac{1}{2}g_{\mu\nu}(\bt)d\theta^{\mu}d\theta^{\nu}, \label{KL2}
\end{eqnarray}
and $g_{\mu\nu}(\bt)$ is the so-called Fisher metric defined by
\begin{eqnarray}
g_{\mu\nu}(\bt)=\sum_{n}\lambda_{n}(\bt)\frac{\partial\gamma_{n}(\bt)}{\partial\theta^{\mu}}\frac{\partial\gamma_{n}(\bt)}{\partial\theta^{\nu}}=\left<\partial_{\mu}\bg\partial_{\nu}\bg\right>. \label{Fisher}
\end{eqnarray}
Roughly speaking, the metric is a kind of two-point correlation function of entropy variation in the information space. The diagonal parts of the metric are all positive, since $g_{\mu\mu}(\bt)=\bigl<(\partial_{\mu}\bg)^{2}\bigr>\ge 0$. Now, we consider whether the information geometrical approach can capture some basic features of the AdS/CFT correspondence, and thus we would like to find the origin of the classical coordinates in the quantum side. The line element squared has Lorentzian signature as well as a warp factor that characterized conformal invariance. Therefore, the positive sign property indicates that we need to take imaginary time when the original quantum system evolves in time and the time coordinate is mapped onto one of $\bt$. Hereafter, we consider the $D$-dimensional information spacetime, and we have
\begin{eqnarray}
g^{\mu\nu}g_{\mu\nu}=D,
\end{eqnarray}
where $g^{\mu\nu}$ is an inverse matrix of $g_{\mu\nu}$.

The Fisher metric has an another form. Let us start with Eq.~(\ref{conservation}), $\left<1\right>=1$. Differentiating both sides of this equation by $\theta^{\nu}$, we obtain
\begin{eqnarray}
\left<\partial_{\nu}\bg\right>=0. \label{gzero}
\end{eqnarray}
One more differentiation by $\theta^{\mu}$ leads to
\begin{eqnarray}
g_{\mu\nu}=\left<(\partial_{\mu}\bg)(\partial_{\nu}\bg)\right>=\left<\partial_{\mu}\partial_{\nu}\bg\right>. \label{Fisher2}
\end{eqnarray}
Thus we have two different representations of the Fisher metric. It should be noted that these two before taking the statistical avarage are not the same
\begin{eqnarray}
\partial_{\mu}\partial_{\nu}\bg\ne(\partial_{\mu}\bg)(\partial_{\nu}\bg).
\end{eqnarray}
This means that the classical spacetime is emerged from averaging procedures of quantum-mechanically fluctuating states.

\section{Hessian Geometry and its Application to Entanglement Spectra}

\subsection{Exponential Family Form of Probability Distribution}

When we start with general definition of the Fisher metric in Eq.~(\ref{Fisher2}), the corresponding geometry is quite complicated and it is hard to extract some physical information. Thus, it is necessary to specify our target model. However, the geometry becomes simple for a particular choice of the distribution function. That is the so-called exponential family form. As is well known, the resulting classical spacetime is described by the Hessian geometry, not the general Riemaniann geometry~\cite{Shima}.

The exponential family form is represented by the following probability distribution
\begin{eqnarray}
\lambda_{n}(\bt) = e^{-\gamma_{n}} = \exp\left\{\theta^{\alpha}F_{n\alpha}-\psi(\bt)\right\}, \label{efamily}
\end{eqnarray}
where $\bt$ is called canonical (natural) parameter. In this definition, $F$ only depends on the index $n$, and $\psi$ only depends on $\bt$. The exponential family covers very wide classes of probability distributions. As already mentioned, the Boltzmann distribution is typical one for this class~\cite{Ruppeiner,Brody}. The distribution is given by $\lambda_{n}(\bt)=e^{\beta E_{n}}/Z$ with inverse temperature $\beta$, and for example $\beta$ normalized by the interaction parameter and $\beta\mu$ with the chemical potential $\mu$ are mapped onto two classical coordinates in the information space. The function $\psi(\bt)$ is given by $\psi(\bt)=\ln Z$, and corresponds to a potential function. In the next subsection, we will discuss applicability of this distribution to the physics of quantum entanglement.

\subsection{Entanglement Spectra for Free Fermions}

\subsubsection{Model}

An important question is whether the entanglement spectrum $\lambda_{n}(\bt)$ is a member of the exponential family or not. To characterize the entanglement by the reduced density matrix, we truncate environmental degrees of freedom. This procedure provides mixed-state feature, and we can expect that this would lead to the exponential family form. In the present stage, we do not have quite general solution for this problem, but in some specific cases we can actually find positive answers for this question. Hereafter, we consider 1D free fermions (or equivalently XY model owing to the presence of Jordan-Wigner transformation). Since the 1D fermion model is typical conformally-invariant one, we may generalize the following approach in terms of CFT.

Let us more precisely formulate our model. We start with a quantum state $\left|\psi\right>$ defined on $(1+1)$-dimensional flat Minkowski spacetime $\mathbb{R}^{1,1}$. We devide the whole system into two spatial regions $A$ and $\bar{A}$, and denote the size of subsystem $A$ as $L$. The Hamiltonian of the whole system $A+\bar{A}$ is defined by
\begin{eqnarray}
H_{A+\bar{A}}=-\sum_{j}\left(c_{j}^{\dagger}c_{j+1}+c^{\dagger}_{j+1}c_{j}\right),
\end{eqnarray}
where $c_{j}^{\dagger}$ and $c_{j}$ are spinless fermion's creation and annihilation operators at site $j$, respectively. We suppose that our quantum state $\left|\psi\right>$ is in the ground state of this Hamiltonian, $H_{A+\bar{A}}\left|\psi\right>=E_{0}\left|\psi\right>$ with the ground-state energy $E_{0}$, or a time-dependent state at time $t$ with this Hamiltonian after some perturbation $O$ to a quantum state $\left|\psi_{0}\right>$, i.e. $\left|\psi\right>=e^{iH_{A+\bar{A}}t}O\left|\psi_{0}\right>$. According to the deomposition of the whole system into two spatial regions, $A$ and $\bar{A}$, we represent $\left|\psi\right>$ by the Schmidt decomposition as
\begin{eqnarray}
\left|\psi\right>=\sum_{n}\sqrt{\lambda_{n}}\left|A;n\right>\otimes\left|\bar{A};n\right>,
\label{wf}
\end{eqnarray}
where $\{\left|A;n\right>\}$ and $\{\left|\bar{A};n\right>\}$ are the Schmidt bases for two subsystems $A$ and $\bar{A}$, respectively. We normalize the Schmidt coefficients or the singular values $\{\lambda_{n}\}$ so that $\left|\psi\right>$ is normalized as
\begin{eqnarray}
\langle\psi|\psi\rangle=\sum_{n}\lambda_{n}=1. \label{conservation}
\end{eqnarray}
The density matrix of the whole system is defined by $\rho_{A+\bar{A}}=\left|\psi\right>\left<\psi\right|$. The reduced density matrix for subsystem $A$ is represented by tracing over environmental degrees of freedom as
\begin{eqnarray}
\rho_{A} = {\rm tr}_{\bar{A}}\left|\psi\right>\left<\psi\right| = \sum_{n}\lambda_{n}\left|A;n\right>\left<A;n\right| .
\end{eqnarray}
Thus, the square of the Schmidt coefficient is the eigenvalue of the reduced density matrix. Usually, we transform the above equation as
\begin{eqnarray}
\rho_{A} = \sum_{n}e^{-\gamma_{n}}\left|A;n\right>\left<A;n\right| = \frac{1}{Z}e^{-\tilde{H}_{A}},
\end{eqnarray}
where $\tilde{H}_{A}=-\theta^{\alpha}F_{\alpha}$ is the so-called entanglement Hamiltonian or modular Hamiltonian (normalized by the entanglement temperature), $\tilde{H}_{A}\left|A;n\right>=-\theta^{\alpha}F_{n\alpha}\left|A;n\right>$, and $\psi(\bt)=\ln Z$. As we have already mentioned, this formally coinsides with the thermal distribution form.

\subsubsection{Mapping from Original Model Parameters to Canonical Parameters}

Our proposal is that the entanglement spectrum $\lambda_{n}(\bt)$ for free fermions actually has the exponential family form. It is clear that the spectrum is a function of subsystem size $L$, filling fraction $\bar{n}$, and time $t$ if the system evolves in time. However, a question is how to determine the correspondence between canonical parameters $\bt$ and the original model parameters $(L,\bar{n},t)$, i.e.
\begin{eqnarray}
\theta^{i}=\theta^{i}(L,\bar{n},t) \; , \; i=1,2,3.
\end{eqnarray}
It is important to notice that this is highly nontrivial correspondence, because the entanglement Hamiltonian $\tilde{H}_{A}$ for subsystem $A$ appears as a result of truncation of $\bar{A}$. In general, the truncation process induces complicated nonlocal interactions.

In the free fermion case, we are going to propose that the canonical parameters should be given by
\begin{eqnarray}
\left(\theta^{1},\theta^{2},\theta^{3}\right)\sim\left(\left(\frac{1}{L}\right)^{2},\frac{1}{L}f^{\prime\prime}(\bar{n},0),\frac{t}{t_{0}}\right). \label{t123}
\end{eqnarray}
where $f=f(\bar{n},x)$ is a scaling function with $x=(l-l_{F})/L$ and $l_{F}$ is a pseudo-Fermi level which will be explained in detail. The prime index in $f$ represents derivative by $x$. The parameter $t_{0}$ is a unit of time that will be determined. The parameter $\theta^{1}$ corresponds to the radial axis of the hyperbolic space (AdS with Euclidean signature).

This nontrivial mapping is crucial to examine emergent geometry from entanglement spectrum. The main purpose in half of this paper is to prove this property.

\subsubsection{Product Form of Reduced Density Matrix for Free Fermions and Pseudo-Fermi Level}

We explain why Eq.~(\ref{t123}) is a reasonable choice. According to Refs.~\cite{Vidal2,Peschel1,Peschel2,Peschel3,Cheong1,Cheong2,Lefevre,Jin}, the reduced density matrix for free fermions is factorized as
\begin{eqnarray}
\rho_{A}=\bigotimes_{l=1}^{L}\varrho_{l}\propto\exp\left\{-\sum_{l=1}^{L}\varphi_{l}n_{l}\right\}, \label{varrho}
\end{eqnarray}
where $\varrho_{l}$ denotes the mixed state of mode $l$, $n_{l}=f_{l}^{\dagger}f_{l}$ is a number operator of a fermion at mode $l$, and $\varphi_{l}$ is its energy (this fermion is different from that in the original model). The operator $\tilde{H}_{A}=\sum_{l=1}^{L}\varphi_{l}n_{l}$ is nothing but the entanglement Hamiltonian. In Refs.~\cite{Cheong1,Cheong2}, $\tilde{H}$ is called as pseudo-Hamiltonian. The entanglement entropy is given by a simple sum of binary entropy of each mode \begin{eqnarray}
S_{A}=\sum_{l=1}^{L}B(\nu_{l}), \label{SL}
\end{eqnarray}
where we take $B(x)=-x\ln x-(1-x)\ln(1-x)$ and $\nu_{l}=\left(e^{\varphi_{l}}+1\right)^{-1}$ is one of normalized eigenvalues of $\varrho_{l}$.

Precise numerical simulation has been performed to examine $L$ and filling dependence on the eigenvalues of $\rho_{L}$~\cite{Peschel1,Cheong2}. The numerical results with fermion filling $\bar{n}$ suggest the presence of a scaling function $f$ defined as
\begin{eqnarray}
\varphi_{l}\left(L,\bar{n}\right)=Lf\left(\bar{n},x\right),
\end{eqnarray}
where $x$ and $l_{F}$ are defined as
\begin{eqnarray}
x=\frac{l-l_{F}}{L},
\end{eqnarray}
and
\begin{eqnarray}
l_{F}=\bar{n}L+\frac{1}{2}.
\end{eqnarray}
The parameter $l_{F}$ is analogous to the Fermi level, and is called as pseudo-Fermi level. To find the largest eigenvalue of $\rho_{A}$ (the minimal energy state), $\lambda_{1}$, we can consider a situation in which the pseudo-energy level $l$ is completely occupied up to the pseudo-Fermi level $l_{F}$. The basic properties of the scaling function are
\begin{eqnarray}
\left\{
\begin{array}{@{\,}l}
\displaystyle
f\left(\bar{n},0\right)=0 \\
\displaystyle
f^{\prime}\left(\bar{n},0\right)>0 \\
\displaystyle
f\left(\bar{n},-x\right)=-f\left(1-\bar{n},x\right)
\end{array}
\right.
.
\end{eqnarray}
At half-filling $\bar{n}=1/2$, $f$ is an odd function, and this symmetry is violated away from half-filling. The particle-hole symmetry is more important here. By using these relations, we can evaluate the dependence of $L$ and $\bar{n}$ on the eigenvalue spectra $\lambda_{n}$ for the full reduced density matrix $\rho_{A}$. 

\subsubsection{On Finding Canonical Parameters}

We consider the difference between the first and second largest eigenvalues. Their corresponding spectra are respectively given by
\begin{eqnarray}
\left\{
\begin{array}{@{\,}l}
\displaystyle
\gamma_{1}=\sum_{l=1}^{l_{F}}\varphi_{l}(L,\bar{n}) \\
\displaystyle
\gamma_{2}=\sum_{l=1}^{l_{F}-1}\varphi_{l}(L,\bar{n}) + \varphi_{l_{F}+1}(L,\bar{n}) 
\end{array}
\right.
,
\end{eqnarray}
and the difference $\Delta=\gamma_{2}-\gamma_{1}$ is evaluated as
\begin{eqnarray}
\Delta = \varphi_{l_{F}+1}(L,\bar{n})-\varphi_{l_{F}}(L,\bar{n}) = Lf\left(\bar{n},1/L\right) . \label{difg}
\end{eqnarray}
Accodring to Eq.~(\ref{efamily}), taking the difference removes contribution from the Hessian potential, and then
\begin{eqnarray}
\Delta=\theta^{\alpha}\left(F_{1\alpha}-F_{2\alpha}\right).
\end{eqnarray}
Note that we can generalize the right hand side by adding a dummy index $\alpha=0$ so that $\theta^{\alpha}F_{n\alpha}=\theta^{0}F_{n0}+\sum_{\alpha=1}^{D}\theta^{\alpha}F_{n\alpha}$ with $\theta^{0}=1$. In this case, Eq.~(\ref{difg}) is expanded as
\begin{eqnarray}
\Delta = f^{\prime}(\bar{n},0) + \frac{1}{2}f^{\prime\prime}(\bar{n},0)\frac{1}{L} + \frac{1}{6}f^{\prime\prime\prime}(\bar{n},0)\left(\frac{1}{L}\right)^{2} + \cdots .
\end{eqnarray}
Here the second derivative $f^{\prime\prime}(\bar{n},0)$ is zero at half-filling $\bar{n}=1/2$. This quantity changes its sign at $\bar{n}=1/2$, and later we will see that this sign change is mapped onto a property of space coordinate in the classical side. Furthermore, the numerical data suggest only weak $\bar{n}$ dependence on $f^{\prime}(\bar{n},0)$ and $f^{\prime\prime\prime}(\bar{n},0)$. Thus, they can approximately be parts of $F_{n\alpha}$. Therefore, we identify $\theta^{1}$ with $L^{-2}(>0)$ and $\theta^{2}$ with $L^{-1}f^{\prime\prime}(\bar{n},0)$, respectively. This is the reason why we select Eq.~(\ref{t123}). For comparison, we show the explicit form of $\gamma_{1}$ as
\begin{eqnarray}
\gamma_{1} &=& \sum_{l=1}^{l_{F}}Lf\left(\bar{n},x\right) \nonumber \\
&=& L\sum_{l=1}^{l_{F}}\left(f^{\prime}\left(\bar{n},0\right)x +\frac{1}{2}f^{\prime\prime}\left(\bar{n},0\right)x^{2}+\cdots \right) \nonumber \\
&=& \frac{1}{2}f^{\prime}\left(\bar{n},0\right)l_{F}\left(1-l_{F}\right) \nonumber \\
&& + \frac{1}{12}f^{\prime\prime}\left(\bar{n},0\right)\frac{l_{F}}{L}\left(2l_{F}^{2}+2l_{F}+1\right) \nonumber \\
&& + \frac{1}{24}f^{\prime\prime\prime}\left(\bar{n},0\right)\left(\frac{l_{F}}{L}\right)^{2}\left(-l_{F}^{2}+2l_{F}-1\right) + \cdots.
\end{eqnarray}

We can also consider the time evolution of the reduced density matrix after some perturbation $O$ by taking
\begin{eqnarray}
\rho_{A} &=& {\rm tr}_{\bar{A}}\left\{ e^{iH_{A+\bar{A}}t}O\left|\psi_{0}\right>\left<\psi_{0}\right|O^{\dagger}e^{-iH_{A+\bar{A}}t} \right\} \nonumber \\
&=& \sum_{m,n}e^{i\left(\epsilon_{m}-\epsilon_{n}\right)t}O_{m}O_{n}^{\ast}{\rm tr}_{\bar{A}}\left|m\right>\left<n\right|,
\end{eqnarray}
where $O_{m}=\left<m\right|O\left|\psi_{0}\right>$ and $H_{A+\bar{A}}\left|m\right>=\epsilon_{m}\left|m\right>$. This is the exponential form and we can identify $\theta^{3}$ with $t$, if the partial density operator ${\rm tr}_{\bar{A}}\left|m\right>\left<n\right|$ is factorized by the tensor product so that each sector has a fixed energy difference $\epsilon_{m}-\epsilon_{n}$~\cite{Nezhadhaghighi}.

It should be noted that the set $\bt$ is mapped onto classical coordinates in the present approach. This is somehow different from the standard AdS/CFT arguments, although the radial axis naturally appears for the construction of the classical geometry. In the standard AdS/CFT, the spacetime coordinates are basically maintained, and we add the radial axis for holographic renormalization. In the present case, however, the theory requires more radical mapping in some sense. This indicates that the AdS/CFT is a quite special case of more general gauge/gravity correspondence.

It is also meaningful to address the concept of bulk/boundary correspondence. The readers may think that this concept is not composed of information geometry. However, if we take large-$L$ limit, then we obtain $\theta^{1}\rightarrow 0$. Later we will see that $\theta^{1}$ is related to the radial axis. These results mean that we approach the boundary of the AdS spacetime when the subsystem size becomes maximum. There is a situation similar to AdS/CFT in which the UV limit of a quantum system is located at the boundary of the AdS spacetime.

Finally, we comment that the $L$ dependence on the entanglement spectra in 1D has been examined in terms of CFT. By using results in Ref.~\cite{Lefevre}, we obtain the following logarithmic Gaussian form for large $n$
\begin{eqnarray}
\lambda_{n} \sim \exp\left\{ \frac{1}{2S}\left(\ln n\right)^{2} - \frac{1}{2}S \right\}.
\end{eqnarray}
If we remenber the logarithmic formula of the entanglement entropy, we see that $\theta^{1}$ is a decreasing function of $L$. Although the functional form is different from our proposal $\theta^{1}\sim L^{-2}$, this might be owing to some approximations for the derivation of these results.

\subsection{Holographic Entanglement Spectra}

The tensor product factorization has been also seen in the singular value decomposition (SVD) of 2D fractal images, and it has been shown that the singular value spectra coinside with the entanglement spectra for 1D free fermions~\cite{Matsueda3,Matsueda4}. In that sense, the SVD spectra for self-similar images can be viewed as holographic entanglement spectra. Here, we holographically interprete the results in the previous subsection.

Consider a completely self-similar image. The image is characterized by a matrix $M$, and each matrix element denotes luminance of each pixel. The matrix data $M$ are constructed by a tensor product of $L$ copies of a $h\times h$ unit cell $H$ owing to the self-similarity~\cite{Matsueda3,Matsueda4}
\begin{eqnarray}
M=\bigotimes_{m=1}^{L}H. \label{snapshot}
\end{eqnarray}
The parameter $L$ is the fractal level, and the $L$-times tensor product creates hierarchy of $L$ different length scales. To evaluate the matrix data in terms of information entropy, we apply SVD to $M$ as
\begin{eqnarray}
M(x,y)=\sum_{n=1}^{h^{L}}U_{n}(x)\sqrt{\Lambda_{n}}V_{n}(y), \label{SVD}
\end{eqnarray}
where $\sqrt{\Lambda_{n}}$ is the singular value spectrum, and $U_{n}(x)$ and $V_{n}(y)$ are column unitary matrices. After normalization of $\Lambda_{n}$ as $\lambda_{n}$, the information entropy, the so-called snapshot entropy, is defined by $S_{snapshot}=-\sum_{n}\lambda_{n}\ln\lambda_{n}$. The correspondence between Eqs.~(\ref{varrho}) and (\ref{snapshot}) exists, when $H$ has two nonzero singular values. The two values correspond to occupation and absence of a fermion at each mode in Eq.~(\ref{varrho}). We take the normalized eigenvalues of $H^{2}$ as $\gamma_{+}$ and $\gamma_{-}$ ($\gamma_{+}+\gamma_{-}=1$). Then, all possible eigenvalues of $M^{2}$ are exactly given by 
\begin{eqnarray}
\lambda_{k}=\gamma_{+}^{k}\gamma_{-}^{L-k}=\gamma_{+}^{k}\left(1-\gamma_{+}\right)^{L-k}, \label{lambda}
\end{eqnarray}
where the label $k$ runs from $0$ to $L$ with degeneracy ${}_{L}C_{k}$, and is essentially equal to the index $n$ in Eqs.~(\ref{efamily}) and (\ref{SVD}). In Eq.~(\ref{SVD}), we have $2^{L}$ nonzero eigenvalues. This number is represented as $2^{L}=(1+1)^{L}=\sum_{k=0}^{L}{}_{L}C_{k}$. For Eq.~(\ref{lambda}), the snapshot entropy is given by
\begin{eqnarray}
S_{snapshot} = -\sum_{k=0}^{L}{}_{L}C_{k}\lambda_{k}\ln\lambda_{k} = B(\gamma_{+})L, 
\end{eqnarray}
which is consistent with $S_{L}$ in Eq.~(\ref{SL}) for free fermions except for the mode dependence ($\nu_{m}\rightarrow\gamma_{+}$).

According to Eq.~(\ref{efamily}), the eigenvalue spectrum is represented as
\begin{eqnarray}
\gamma_{k}=-\ln\left(\gamma_{+}^{k}\gamma_{-}^{L-k}\right).
\end{eqnarray}
We calculate the difference between $\gamma_{L}$ and $\gamma_{L-1}$, and then we obtain
\begin{eqnarray}
\gamma_{L}-\gamma_{L-1}=\ln\left(\frac{\gamma_{-}}{\gamma_{+}}\right)=N.
\end{eqnarray}
This is a canonical parameter. The ratio $\gamma_{-}/\gamma_{+}$ controls luminance in the image. We find $\gamma_{+}=f(N)$ with the Fermi distribution function $f(x)=(e^{x}+1)^{-1}$. Thus, this is closely related to $\bar{n}$ in the quantum case. In the fractal case, there is only one relavant parameter, and the system size $L$ does not appear. For a small-$L$ region, the self-similar structure of a fractal image is smeared out. Thus, the exact scale invariance appears in the large-$L$ limit. That would be the reason why we only look at one relevant parameter.

\subsection{Hessian Geometry: Close Relationship Among Hessian Potential, Entanglement Entropy, and Fisher Metric}

Let us move to the geometrical analysis of information space under the condition of exponential family form. In this subsection, we first summarize various representations of the Fisher metric. In the exponential family, the spectrum is given by
\begin{eqnarray}
\bg_{n}(\bt)=\psi(\bt)-\theta^{\alpha}F_{n\alpha}. \label{gamma2}
\end{eqnarray}
The first and second derivatives of $\bg$ are respectively given by
\begin{eqnarray}
\partial_{\mu}\bg = \partial_{\mu}\psi(\bt)-F_{n\mu}, \label{partial1}
\end{eqnarray}
and
\begin{eqnarray}
\partial_{\mu}\partial_{\nu}\bg = \partial_{\mu}\partial_{\nu}\psi(\bt). \label{partial2}
\end{eqnarray}
After the second derivative, $\psi(\bt)$ can be identified with $\bg$. According to Eq.~(\ref{Fisher2}), the Fisher metric is represented by $\psi(\bt)$ as
\begin{eqnarray}
g_{\mu\nu} = \left<\partial_{\mu}\partial_{\nu}\bg\right>=\partial_{\mu}\partial_{\nu}\psi(\bt). \label{rep1}
\end{eqnarray}
Thus we need not to take the statistical average in the last equation, since $\psi(\bt)$ does not depend on the stochastic index $n$. This is the so-called Hessian structure, and $\psi(\bt)$ is a Hessian potential~\cite{Shima}. This is a real version of the K\"{a}hler potential in the complex manifold theory. This simplification has been well-known, but we will see that this is physically quite important in later discussions.

Taking the average of Eq.~(\ref{partial1}) with the help of Eq.~(\ref{gzero}), we find 
\begin{eqnarray}
\left<F_{\mu}\right>=\partial_{\mu}\psi(\bt)=\eta_{\mu}(\bt). \label{fav}
\end{eqnarray}
This corresponds to Legendre transformation in terms of thermodynamics. By using this equality, we find
\begin{eqnarray}
g_{\mu\nu} = \left<\partial_{\mu}\bg\partial_{\nu}\bg\right> = \left<F_{\mu}F_{\nu}\right>-\left<F_{\mu}\right>\left<F_{\nu}\right>. \label{rep2}
\end{eqnarray}
Thus, the metric is a covariance matrix of $F_{\mu}$.

These results are also derived from normalization condition of the entanglement spectrum in Eq.~(\ref{efamily}). Let us start with the following relation:
\begin{eqnarray}
1=\sum_{n}\lambda_{n}(\bt)=e^{-\psi}\sum_{n}e^{\theta^{\alpha}F_{n\alpha}}.
\end{eqnarray}
By solving this for $\psi(\bt)$, we obtain
\begin{eqnarray}
\psi(\bt)=\ln\left\{ \sum_{n}e^{\theta^{\alpha}F_{n\alpha}} \right\}, \label{ZEE}
\end{eqnarray}
and the first and second derivatives are respectively given by
\begin{eqnarray}
\partial_{\nu}\psi(\bt)=\frac{\sum_{n}F_{n\nu}e^{\theta^{\alpha}F_{n\alpha}}}{\sum_{n}e^{\theta^{\alpha}F_{n\alpha}}}=\frac{\sum_{n}F_{n\nu}\lambda_{n}}{\sum_{n}\lambda_{n}}=\left<F_{\nu}\right>,
\end{eqnarray}
and $g_{\mu\nu}=\partial_{\mu}\partial_{\nu}\psi=\left<F_{\mu}F_{\nu}\right>-\left<F_{\mu}\right>\left<F_{\nu}\right>$. The function inside of logarithm in Eq.~(\ref{ZEE}) is a kind of partition function. We call this as entanglement partition function, and thus $\psi$ can be regarded as entanglement free energy normalized by entanglement temperature.

Next we consider the information entropy. Since we are going to treat the entanglement spectrum, the entropy given by this spectrum is of course the entanglement entropy. Taking the statistical avarage of Eq.~(\ref{gamma2}), we find
\begin{eqnarray}
S(\bt)=\psi(\bt)-\theta^{\alpha}\left<F_{\alpha}\right>. \label{sav}
\end{eqnarray}
Combining Eq.~(\ref{sav}) with Eq.~(\ref{fav}), we can rewrite the entropy as
\begin{eqnarray}
S(\bt) &=& \psi(\bt)-\theta^{\alpha}\partial_{\alpha}\psi(\bt). \label{solve}
\end{eqnarray}
Since $\psi$ is the entanglement free energy, this represents a generalized first law of thermodynamics. Here, we do not go into detailed examination of general properties of this law, but it would be interesting to compare this with recent works on entanglement thermodynamics~\cite{Blanco,Casini,Wong,Takayanagi2,Takayanagi3,Faulkner,Banerjee,Nima}.

The first derivative of the entropy is represented as
\begin{eqnarray}
\partial_{\nu}S(\bt)=-\theta^{\alpha}\partial_{\alpha}\partial_{\nu}\psi(\bt)=-\theta^{\alpha}g_{\alpha\nu}, \label{dS}
\end{eqnarray}
and the second derivative is also represented as
\begin{eqnarray}
\partial_{\mu}\partial_{\nu}S(\bt)=-g_{\mu\nu}(\bt)+\theta^{\alpha}T_{\alpha\mu\nu}(\bt), \label{sd1}
\end{eqnarray}
where
\begin{eqnarray}
T_{\alpha\mu\nu}=-\partial_{\alpha}\partial_{\mu}\partial_{\nu}\psi(\bt). \label{TTT}
\end{eqnarray}
By using Eq.~(\ref{dS}), we find
\begin{eqnarray}
(\partial_{\mu}S)(\partial_{\nu}S) = \theta^{\alpha}\theta^{\beta}g_{\mu\alpha}g_{\nu\beta}.
\end{eqnarray}
Operating $g^{\mu\nu}$ and taking convention for indices $\mu$ and $\nu$, we obtain
\begin{eqnarray}
g^{\mu\nu}(\partial_{\mu}S)(\partial_{\nu}S) = \theta^{\alpha}\theta^{\beta}g_{\alpha\beta}. \label{diffeq2}
\end{eqnarray}

\subsection{Mapping of CFT Data onto AdS Metric}

\subsubsection{Outline of Mapping}

Let us look at how CFT data are mapped onto hyperbolic (Euclidean AdS) metric. At first, we present a simple outline to facilitate better understanding of this mapping. Going back to Eq.~(\ref{solve}), we consider a case that
\begin{eqnarray}
S(\theta^{1})=-\kappa\ln\theta^{1}= +2\kappa\ln L,
\end{eqnarray} according to the logarithmic formula of the entanglement entropy for 1D critical systems~\cite{Holzhey,Calabrese1,Calabrese2,Calabrese3,AAAL,Liu,Nezhadhaghighi,Hubeny}. In this case, the solution of this differential equation is
\begin{eqnarray}
\psi(\theta^{1})=S(\theta^{1})-\kappa+F_{1}\theta^{1}, \label{outline}
\end{eqnarray}
with an arbitrary constant $F_{1}$. Thus, the potential can be identified with the entropy after the second derivative
\begin{eqnarray}
g_{11}=\partial_{1}\partial_{1}\psi(\theta^{1})=\partial_{1}\partial_{1}S(\theta^{1}).
\end{eqnarray}
The third term $F_{1}\theta^{1}$ is absorbed into $\theta^{\alpha}F_{n\alpha}$. Be careful about a fact that the sign of the first term in Eq.~(\ref{sd1}) is opposite. According to Eq.~(\ref{rep1}), the metric is given by
\begin{eqnarray}
g_{11}=\left(\theta^{1}\right)^{-2},
\end{eqnarray}
which is nothing but the warp factor of hyperbolic space. Although the above approach does not consider the other components $\theta^{2}$ and $\theta^{3}$, we expect that the approach captures some essential features of information-gemetrical representation of the AdS/CFT correspondence.

In the next two subsections, we examine the exact representation of the mapping from the Hessian potential (or entanglement entropy) in a quantum system onto the Fisher metric. For this purpose, we propose the following Hessian potential
\begin{eqnarray}
\psi(\bt)=-\kappa\ln f=-\kappa\ln\left\{ \theta^{1}-\frac{1}{2}\sum_{i=2}^{D}\left(\theta^{i}\right)^{2} \right\}, \label{fullpotential}
\end{eqnarray}
with a positive constant $\kappa$ corresponding to the central charge in the quantum side. On the other hand, we will see that $-1/4\kappa$ corresponds to the sectional curvature in the classical side. Equation~(\ref{fullpotential}) is direct extention of Eq.~(\ref{outline}) except for some irrelevant terms, and the extra factor $\sum_{i=2}^{D}(\theta^{i})^{2}$ is introduced. We realize the presence of the bilinear form
\begin{eqnarray}
\xi(\theta^{2},..,\theta^{D})=\frac{1}{2}\sum_{i=2}^{D}(\theta^{i})^{2},
\end{eqnarray}
since $\theta^{1}=L^{-2}$ against $\theta^{2}\propto L^{-1}$ in the free fermion case. In that sense, this derivation is not so highly nontrivial. When we take $(L,\bar{n},t)$-representation, we prove that this potential $\psi(\theta^{1}(L,\bar{n},t),\theta^{2}(L,\bar{n},t),...,\theta^{D}(L,\bar{n},t))$ is identical to the entanglement entropy $S(L,\bar{n},t)$ in free fermions. On the other hand, when we calculate $g_{\mu\nu}=\partial_{\mu}\partial_{\nu}\psi$ according to Eq.~(\ref{rep1}), we show that this finally leads to hyperbolic metric.

\subsubsection{Quantum Side}

Let us first consider the quantum side. According to Eq.~(\ref{fullpotential}), a parameter region we consider should be
\begin{eqnarray}
\theta^{1}>(1/2)\sum_{i=2}^{D}(\theta^{i})^{2}.
\end{eqnarray}
We would like to confirm that this assumption is reasonable for our choice of the parameters in Eq.~(\ref{t123}). Substituting Eq.~(\ref{t123}) into this inequality, we obtain
\begin{eqnarray}
\left(\frac{1}{L}\right)^{2} > \frac{1}{2}\left\{ \left(\frac{f^{\prime\prime}(\bar{n},0)}{L}\right)^{2}+\left(\frac{t}{t_{0}}\right)^{2} \right\}.
\end{eqnarray}
This condition is satisfied, when we take $t_{0}\sim L$ and $t<t_{0}$. We take $t_{0}=L/m$. The time scale $t_{0}\sim L$ indicates time of fermion motion from one side to the other in a finite size system with length $L$. Soon after, we will discuss dynamics of the entanglement entropy after quantum quench, and in this case we actualy find  that the conditions are reasonable.

As we have already mentioned, in the large $L$-limit, the potential function in Eq.~(\ref{fullpotential}) is equivalent to the logarithmic entropy formula in CFT. Let us examine a role of the other components on the scaling formula of the entanglement entropy. We expand the potential as
\begin{eqnarray}
\psi(\bt) &=& -\kappa\ln\left\{ \theta^{1}-\frac{1}{2}\sum_{i=2}^{D}\left(\theta^{i}\right)^{2} \right\} \nonumber \\
&=& -\kappa\ln\theta^{1}-\kappa\ln\left\{ 1-\frac{1}{2}\sum_{i=2}^{D}\frac{\left(\theta^{i}\right)^{2}}{\theta^{1}} \right\} \nonumber \\
&\simeq& -\kappa\ln\theta^{1} + \frac{1}{2}\kappa\sum_{i=2}^{D}\frac{\left(\theta^{i}\right)^{2}}{\theta^{1}}.
\end{eqnarray}
Substituting Eq.~(\ref{t123}) into this result, we find
\begin{eqnarray}
\psi(L,\bar{n},t) \simeq 2\kappa\ln L + \frac{1}{2}\kappa\left\{ \left(f^{\prime\prime}(\bar{n},0)\right)^{2} + m^{2}t^{2} \right\}.
\end{eqnarray}
Since we can identify $\psi(L,\bar{n},t)$ with $S(L,\bar{n},t)$, this is actually consistent with the entanglement entropy scaling for $\kappa=c/6$ with the central charge $c$~\cite{Holzhey,Calabrese1,Calabrese2,Calabrese3,AAAL,Liu,Nezhadhaghighi,Hubeny}.
In particular, the quadratic time-dependent feature of the entropy has been found recently for a time region $t\ll L$~\cite{Nezhadhaghighi}. 
Furthermore, the coefficient $\kappa$ contains information of the central charge as we have already mentioned. This feature is also consistent with recent numerical simulation in which the coefficients of time-dependent terms also depend on $c$~\cite{Nezhadhaghighi}. The filling dependence is not still well-known, but a related description appears in Ref.~\cite{AAAL}.

\subsubsection{Classical Side}

For the full potential function $\psi(\bt)$ in Eq.~(\ref{fullpotential}), there exists a parameter set $\by=\left(y^{1},y^{2},...,y^{D}\right)$ for which the metric tensor is exactly the hyperbolic form. For this proof, we first introduce the Legendre transformation $\eta_{i}=\partial_{i}\psi$ and
\begin{eqnarray}
\left\{
\begin{array}{@{\,}ll}
\displaystyle
\eta_{1}=-\frac{\kappa}{f} & \\
\displaystyle
\eta_{i}=\frac{\kappa\theta^{i}}{f} & (i=2,3,...,D)
\end{array}
\right.
.
\end{eqnarray}
By using this paramter set conjugate to $\bt$, the metric is represented as
\begin{eqnarray}
g_{\mu\nu}=\partial_{\mu}\partial_{\nu}\psi=\partial_{\mu}\eta_{\nu}=\partial_{\nu}\eta_{\mu}.
\end{eqnarray}
The new parameter set $\by$ is defined by
\begin{eqnarray}
\left\{
\begin{array}{@{\,}ll}
\displaystyle
y^{1}=\sqrt{f} & \\
\displaystyle
y^{i}=\sqrt{\kappa}\theta^{i} & (i=2,3,...,D)
\end{array}
\right.
. \label{yy}
\end{eqnarray}
These parameter regions are given by
\begin{eqnarray}
y^{1}>0 \; , \; -\infty<y^{i}<\infty.
\end{eqnarray}
Note that in this coordinate the Hessian potential is represented as $\psi(\by)=-2\kappa\ln y^{1}$. This means that the radial axis characterizes the magnitude of the entanglmenent entropy, and thus the entropy is a key factor of characterizing holographic renormalization.

Then, $\bt$ and its conjugate $\mbox{\boldmath $\eta$}$ are represented by $\by$ as
\begin{eqnarray}
\left\{
\begin{array}{@{\,}ll}
\displaystyle
\theta^{1}=(y^{1})^{2}+\frac{1}{2\kappa}\sum_{i=2}^{D}(y^{i})^{2} & \\
\displaystyle
\theta^{i}=\frac{1}{\sqrt{\kappa}}y^{i} & (i=2,3,...,D)
\end{array}
\right.
,
\end{eqnarray}
and
\begin{eqnarray}
\left\{
\begin{array}{@{\,}ll}
\displaystyle
\eta_{1}=-\frac{\kappa}{(y^{1})^{2}} & \\
\displaystyle
\eta_{i}=\frac{\sqrt{\kappa}y^{i}}{(y^{1})^{2}} & (i=2,3,...,D)
\end{array}
\right.
.
\end{eqnarray}
The metric tensor is transformed into
\begin{eqnarray}
g &=& \left(\partial_{\nu}\eta_{1}\right)d\theta^{\nu}d\theta^{1} + \sum_{i=2}^{D}\left(\partial_{\nu}\eta_{i}\right)d\theta^{\nu}d\theta^{i} \nonumber \\
&=& d\eta_{1}d\theta^{1} + \sum_{i=2}^{D}d\eta_{i}d\theta^{i} \nonumber \\
&=& \frac{1}{(y^{1})^{2}}\left\{ 4\kappa(dy^{1})^{2} + \sum_{i=2}^{D}(dy^{i})^{2} \right\}.
\end{eqnarray}
We have arrived at the hyperbolic metric with the sectional curvature $-1/4\kappa$. Therefore, the entanglement entropy or the Hessian potential of free fermions is mapped onto the hyperbolic metric exactly. In this sense, the Fisher geometry can capture basic properties of the AdS/CFT correspondences. Again note that we should take Euclidean signature for the time coordinate, and this situation is quite similar to the entropy calculation of the black hole. In our case, the time evolution originates in the original quantum system and then the imaginary factor naturally appear through the time evolution operator. This might resolve a long-standing problem for this sign issue.

We finally comment on a physical unit. We change the notation as $y^{i}=2\sqrt{\kappa}\tilde{y}^{i}$, and $g\rightarrow 4\kappa g$. Then, $4\kappa$ can be regarded as the curvature radius $l$ of the hyperbolic space. In the AdS/CFT case, the coefficient is actually related to the curvature radius $l$ of the AdS spacetime which is mapped onto the central charge $c$ by the Brown-Henneaux formula $c=3l/2G$ with the Newton constant $G$~\cite{Brown}. Although there is no gravitational constant in our parameter spacetime, the linear relation of $\kappa$ to $l$ is a strong evidence of the presence of the Brown-Henneaux-type formula.

\subsection{Properties of rank-three tensor $T_{\lambda\mu\nu}$}

To facilitate later discussion, we would like to introduce an another form of the second derivative of the entropy and compare it with Eq.~(\ref{sd1}). We will find the basic properties of $T_{\lambda\mu\nu}$ previously introduced in Eq.~(\ref{TTT}). It should be noted that hereafter we should use the canonical parameters $\bt$ instead of $\by$. Starting with the definition of the entropy in Eq.~(\ref{entropy}), we find
\begin{eqnarray}
\partial_{\nu}S(\bt)=-\sum_{n}\left(\partial_{\nu}\lambda_{n}\right)\ln\lambda_{n}.
\end{eqnarray}
The second derivative of this is given by
\begin{eqnarray}
\partial_{\mu}\partial_{\nu}S(\bt) &=& -g_{\mu\nu}-\sum_{n}\left(\partial_{\mu}\partial_{\nu}\lambda_{n}\right)\ln\lambda_{n}. \label{sd2}
\end{eqnarray}
Comparing this with Eq.~(\ref{sd1}), we find
\begin{eqnarray}
\theta^{\alpha}T_{\alpha\mu\nu}(\bt)=\sum_{n}\left(\partial_{\mu}\partial_{\nu}\lambda_{n}\right)\gamma_{n},
\end{eqnarray}
and then
\begin{eqnarray}
\theta^{\alpha}T_{\alpha\mu\nu}(\bt) = \left<\bg\left(\partial_{\mu}\bg\right)\left(\partial_{\nu}\bg\right)\right>-S(\bt)g_{\mu\nu}. \label{tt1}
\end{eqnarray}
This means higher-order metric correction coupled to fluctuation of information embedded into the parameter space. We will see later that this relation is useful for evaluation of the Ricci tensor.

\subsection{Christoffel Symbol}

The Chistoffel symbol is defined by
\begin{eqnarray}
\Gamma^{\lambda}_{\;\mu\nu}=\frac{1}{2}g^{\lambda\tau}\left(\partial_{\mu}g_{\nu\tau}+\partial_{\nu}g_{\mu\tau}-\partial_{\tau}g_{\mu\nu}\right), \label{Christoffel}
\end{eqnarray}
where $g^{\lambda\tau}$ is the inverse matrix of the Fisher metric. It is quite easy to notice
\begin{eqnarray}
\Gamma^{\lambda}_{\;\mu\nu}=\frac{1}{2}g^{\lambda\tau}\partial_{\tau}\partial_{\mu}\partial_{\nu}\psi(\bt)= -\frac{1}{2}g^{\lambda\tau}T_{\tau\mu\nu}.
\end{eqnarray}
To find an another form, we differentiate $g_{\mu\nu}=\left<\partial_{\mu}\bg\partial_{\nu}\bg\right>$ by $\theta^{\lambda}$. Then we obtain
\begin{eqnarray}
\partial_{\lambda}g_{\mu\nu} &=& -\left<\partial_{\lambda}\bg\partial_{\mu}\bg\partial_{\nu}\bg\right> \nonumber \\
&& +\left<\left(\partial_{\lambda}\partial_{\mu}\bg\right)\partial_{\nu}\bg\right>+\left<\partial_{\mu}\bg\left(\partial_{\lambda}\partial_{\nu}\bg\right)\right> \nonumber \\
&=& -\left<\partial_{\lambda}\bg\partial_{\mu}\bg\partial_{\nu}\bg\right> + g_{\lambda\mu}\left<\partial_{\nu}\bg\right> + g_{\lambda\nu}\left<\partial_{\mu}\bg\right> \nonumber \\
&=& -\left<\partial_{\lambda}\bg\partial_{\mu}\bg\partial_{\nu}\bg\right>,
\end{eqnarray}
and find
\begin{eqnarray}
T_{\lambda\mu\nu}=-\partial_{\lambda}\partial_{\mu}\partial_{\nu}\psi(\bt)=\left<\partial_{\lambda}\bg\partial_{\mu}\bg\partial_{\nu}\bg\right>.
\end{eqnarray}
This tensor plays a central role in the Hessian geometry.

\subsection{Ricci and Einstein Tensors}

The Ricci tensor is given by
\begin{eqnarray}
R_{\mu\nu} &=& R^{\sigma}_{\;\mu\sigma\nu} \nonumber \\
&=& \partial_{\sigma}\Gamma^{\sigma}_{\;\mu\nu}-\partial_{\nu}\Gamma^{\sigma}_{\;\mu\sigma} + \Gamma^{\sigma}_{\;\rho\sigma}\Gamma^{\rho}_{\;\mu\nu} - \Gamma^{\sigma}_{\;\rho\nu}\Gamma^{\rho}_{\;\mu\sigma}. \label{Riemann}
\end{eqnarray}
Here, we can use the following identity
\begin{eqnarray}
\partial_{\sigma}g^{\mu\nu}=-g^{\mu\alpha}g^{\nu\beta}\partial_{\sigma}g_{\alpha\beta}.
\end{eqnarray}
In cases of exponential family, the Ricci tensor is transformed into
\begin{eqnarray}
R_{\mu\nu} &=& \Gamma^{\sigma}_{\;\rho\nu}\Gamma^{\rho}_{\;\mu\sigma} - \Gamma^{\sigma}_{\;\rho\sigma}\Gamma^{\rho}_{\;\mu\nu} \nonumber \\
&=& \frac{1}{4}g^{\sigma\tau}g^{\rho\zeta}\left( T_{\zeta\mu\sigma}T_{\rho\nu\tau}-T_{\rho\sigma\tau}T_{\zeta\mu\nu} \right). \label{Riemann2}
\end{eqnarray}
Thus, the Ricci tensor is basically a bilinear form of the rank-three tensor $T_{\alpha\mu\nu}$.

The Einstein tensor is represented as
\begin{eqnarray}
G_{\mu\nu}=R_{\mu\nu}-\frac{1}{2}g_{\mu\nu}R,
\end{eqnarray}
where the scalar curvature is defined by
\begin{eqnarray}
R=g^{\alpha\beta}R_{\alpha\beta}.
\end{eqnarray}

\section{Information-Geometrical Meaning of Einstein Equation}

\subsection{Evaluation of Rank-Three Tensor $T_{\lambda\mu\nu}$}

Based on the results discussed in the previous section, we would like to examine dynamics of the information spacetime. This examination provides us with information-geometrical interpretation of the Einstein equation. Now we are thinking about both of filling dependence and time evolution in the quantum side, and such dynamical change in the quantum side is mapped onto re-distribution of entropy data embedded into the spacetime by the quite non-uniform manner in general. This situation is quite similar to the presence of the energy-monentum tensor in the standard general relativity. This is the reason why we consider the Einstein equation. We have already taken the Fisher metric as a result of a curved geometry, and thus we are going to evaluate the Einstein tensor. Then, we will obtain the energy-momentum tensor as an origin of the Fisher metric. Of course the origin should be related to the entanglement entropy. Therefore, we conjecture that the entropy behaves as a scaler field to induce the energy-momentum tensor. Later we will confirm this conjecture.

Let us consider the Ricci tensor in Eq.~(\ref{Riemann2}). To treat this quantity, we need to obtain a more convenient form of the rank-three tensor $T_{\lambda\mu\nu}$. For this purpose, we differentiate Eq.~(\ref{tt1}) by $\theta^{\lambda}$. Then we find
\begin{eqnarray}
T_{\lambda\mu\nu}+\theta^{\alpha}\partial_{\lambda}T_{\alpha\mu\nu} &=& -\left<(\partial_{\lambda}\bg)\gamma(\partial_{\mu}\bg)(\partial_{\nu}\bg)\right> + T_{\lambda\mu\nu} \nonumber \\
&& + \left<\bg(\partial_{\lambda}\partial_{\mu}\bg)(\partial_{\nu}\bg)\right> \nonumber \\
&& + \left<\bg(\partial_{\mu}\bg)(\partial_{\lambda}\partial_{\nu}\bg)\right> \nonumber \\
&& - g_{\mu\nu}\partial_{\lambda}S(\bt)-S(\bt)\partial_{\lambda}g_{\mu\nu},
\end{eqnarray}
and this leads to
\begin{eqnarray}
\theta^{\alpha}\partial_{\alpha}T_{\lambda\mu\nu} &=& -\left<\bg(\partial_{\lambda}\bg)(\partial_{\mu}\bg)(\partial_{\nu}\bg)\right> \nonumber \\
&& + g_{\mu\lambda}\left<\bg(\partial_{\nu}\bg)\right> + g_{\nu\lambda}\left<\bg(\partial_{\mu}\bg)\right> \nonumber \\
&& - g_{\mu\nu}\partial_{\lambda}S(\bt) + S(\bt)T_{\lambda\mu\nu}. \label{evaluation}
\end{eqnarray}
Here we have the following relation
\begin{eqnarray}
\left<\bg(\partial_{\mu}\bg)\right>=\sum_{n}(\ln\lambda_{n})\partial_{\mu}\lambda_{n}=-\partial_{\mu}S(\bt),
\end{eqnarray}
and then Eq.~(\ref{evaluation}) is transformed into
\begin{eqnarray}
\theta^{\alpha}\partial_{\alpha}T_{\lambda\mu\nu} &=& S(\bt)T_{\lambda\mu\nu}-\left<\bg(\partial_{\lambda}\bg)(\partial_{\mu}\bg)(\partial_{\nu}\bg)\right> \nonumber \\
&& - g_{\mu\nu}\partial_{\lambda}S(\bt) - g_{\mu\lambda}\partial_{\nu}S(\bt) - g_{\nu\lambda}\partial_{\mu}S(\bt). \nonumber \\ \label{Tbefore}
\end{eqnarray}

Here, we introduce two assumptions. At first, we assume the decompostion as
\begin{eqnarray}
\left<\bg(\partial_{\lambda}\bg)(\partial_{\mu}\bg)(\partial_{\nu}\bg)\right> \simeq S(\bt)T_{\lambda\mu\nu}.
\end{eqnarray}
This assumption would be reasonable for quantum critical systems. The spectrum $\lambda_{n}$ is closely related to correlation function and shows power law decay for the index $n$ at criticality. This means $\lambda_{n}\propto n^{-\Delta}$ with an exponent $\Delta$, and then the spectrum only behaves as logarithmic increase for $n$. Thus the $n$ dependence on the spectrum is not strong. Next we assume
\begin{eqnarray}
\theta^{\alpha}\partial_{\alpha}T_{\lambda\mu\nu}=-AT_{\lambda\mu\nu},
\end{eqnarray}
with a constant factor $A$. The second assumptoin is correct if the metric is a power-law-decay function like hyperbolic (or AdS) metric. For these assumptions, we obtain
\begin{eqnarray}
T_{\lambda\mu\nu}=\frac{1}{A}\left\{ g_{\mu\nu}\partial_{\lambda}S(\bt) + g_{\mu\lambda}\partial_{\nu}S(\bt) + g_{\nu\lambda}\partial_{\mu}S(\bt) \right\}. \label{Tapprox}
\end{eqnarray}

\subsection{Energy-Momentum Tensor in the Information Spacetime}

As we have already examined, the Fisher metric for the entanglement spectrum for free fermions is hyperbolic (or AdS) type. Thus, this is clearly a solution of the vacuum Einstein equation with negative cosmological constant. This proceduce is quite opposite to those in the standard general relativity, since in the present case we just substitute the Fisher metric into the definition of the Einstein tensor. Here, we are very much interested in how the entanglement spectrum away from the conformal fixed point behaves in the classical side. The deformation of the spectrum from those at the critical point induces some dynamics in the information spacetime, and we imagine that the source of the dynamics in the classical side is related to the energy-momentum tensor for a given Lagrangian. We would like to find what is the field strength in the Lagrangian.

Based on the previous subsection, the Ricci tensor is evaluated after some algebrae as
\begin{eqnarray}
R_{\mu\nu} &=& \left(\frac{1}{2A}\right)^{2}(2-D)(\partial_{\mu}S)(\partial_{\nu}S) \nonumber \\
&& - \left(\frac{1}{2A}\right)^{2}Dg_{\mu\nu}(\partial_{\alpha}S)(\partial^{\alpha}S),
\end{eqnarray}
and the Einstein tensor is represented as
\begin{eqnarray}
G_{\mu\nu} &=& \left(\frac{1}{2A}\right)^{2}(2-D)(\partial_{\mu}S)(\partial_{\nu}S) \nonumber \\
&& + \left(\frac{1}{2A}\right)^{2}\frac{(D-2)(D+1)}{2}g_{\mu\nu}(\partial_{\alpha}S)(\partial^{\alpha}S). \label{gmn} \nonumber \\
\end{eqnarray}
We find that the minimal dimension is larger than two to find some nontrivial geometric structure in the information spacetime. Be careful again that this equation is in $\bt$ representation.

Here let us look at free fermion case. Of course, this case exactly satisfies the vacuum Einstein equation in definition. However now we take some approximations to derive Eq.~(\ref{gmn}), and this feature might be violated. Thus, it seems meaningful to confirm reliability of our approximation by looking at whether all the terms in the right hand side in Eq.~(\ref{gmn}) are reduced to a negative cosmological constant in the free fermion case.

According to Eq.~(\ref{fullpotential}), the second term in right hand side of Eq.~(\ref{gmn}) becomes a positive constant:
\begin{eqnarray}
(\partial_{\alpha}S)(\partial^{\alpha}S)\simeq\kappa,
\end{eqnarray}
and contributes to a negative cosmological constant for $D>2$. On the other hand, the first term of Eq.~(\ref{gmn}) for $\mu=\nu=1$ is evaluated as follows:
\begin{eqnarray}
(\partial_{1}S)^{2}\simeq (\partial_{1}\psi)^{2}=\frac{\kappa^{2}}{(\theta^{1}-\xi)^{2}}=\kappa\partial_{1}\partial_{1}\psi.
\end{eqnarray}
For $i=2,3$, we also find
\begin{eqnarray}
(\partial_{i}S)^{2}\simeq (\partial_{i}\psi)^{2}=\frac{(\kappa\theta^{i})^{2}}{(\theta^{1}-\xi)^{2}}\sim\kappa\partial_{i}\partial_{i}\psi.
\end{eqnarray}
This does not show perfect match with $\kappa\partial_{i}\partial_{i}\psi$, but this is not bad at least for $\theta^{1}-(1/2)\xi<(\theta^{i})^{2}$. Thus, the first term of Eq.~(\ref{gmn}) seems to absorbed into the cosmological constant. The constant is defined by
\begin{eqnarray}
\Lambda = -\left(\frac{1}{2A}\right)^{2}\frac{(D-2)(D-1)}{2}\kappa <0.
\end{eqnarray}
Note that this $D$ dependence is familiar in curved geometry, and thus we think that our approximation gives reasonable results when we go away from critical point.

Away from critical point, this feature would be deformed, but we approximately treat the second term as the constant. We introduce the difference of the entanglement entropy from its critical value $S_{0}$ as $\phi=S-S_{0}$. Then, Eq.~(\ref{gmn}) is reduced to the following form
\begin{eqnarray}
G_{\mu\nu}+g_{\mu\nu}\Lambda \simeq \left(\frac{1}{2A}\right)^{2}(2-D)(\partial_{\mu}\phi)(\partial_{\nu}\phi). \label{Einstein1}
\end{eqnarray}

By the above analysis, we notice that $\phi$ behaves as a free scaler field in this classical spacetime. Thus, let us define the effective Lagrangian for the scaler field $\phi$ as
\begin{eqnarray}
L = \frac{1}{2}(\partial_{\alpha}\phi)(\partial^{\alpha}\phi),
\end{eqnarray}
and then the energy-momentum tensor can be represented as
\begin{eqnarray}
T_{\mu\nu} = g_{\mu\lambda}\frac{\partial L}{\partial(\partial_{\lambda}\phi)}\partial_{\nu}\phi - g_{\mu\nu}L,
\end{eqnarray}
and then Eq.~(\ref{Einstein1}) is equal to the Einstein equation
\begin{eqnarray}
G_{\mu\nu}+g_{\mu\nu}\Lambda = a T_{\mu\nu}, \label{Einstein2}
\end{eqnarray}
where
\begin{eqnarray}
a=\frac{1}{2A^{2}}(D-2).
\end{eqnarray}
This is what we would like to obtain here.

There is a comment about approximation to derive Eq.~(\ref{Tapprox}) from Eq.~(\ref{Tbefore}). This leads to the presence of the free scaler field, and more precise treatment might induce some interaction for the field or more complicated fields. This would be a future work.

\section{Summary}

We have examined geometric structure of information spacetime spanned by the Fisher metric for the exponential family class of probability distribution. This class matches with the entanglement spectrum for 1D free fermion system that is a typical CFT, and we can naturally examine the embedding of quantum critical data into classical manifold. This situation is quite similar to the physics of AdS/CFT. We have found that Eqs.~(\ref{rep1}) and (\ref{solve}) are crucial to understand a close relationship among the entanglement entropy, the Hessian potential, and the classical metric. Actually, this relationship leads to the correspondence between AdS and CFT. We have also found that the Fisher metric for free fermions at the critical point is the vacuum solution of the Einstein equation and the deformation of the entanglement spectrum or the entanglement entropy from the critical point produces the energy-momentum tensor. There, the classical scaler field is equal to the entanglement entropy embedded into the spacetime, and the field fluctuation from the uniform hyperbolic background is essential for the emergence of the energy-momentum tensor. Therefore, the Einstein equation describing the dynamics of the information spacetime corresponds to a kind of the equation of states in the quantum side. This interpretation seems to be consistent with the entanglement thermodynamics~\cite{Blanco,Casini,Wong,Takayanagi2,Takayanagi3,Faulkner,Banerjee,Nima}.

\begin{figure}[htbp]
\vspace{2mm}
\begin{center}
\includegraphics[width=8.5cm]{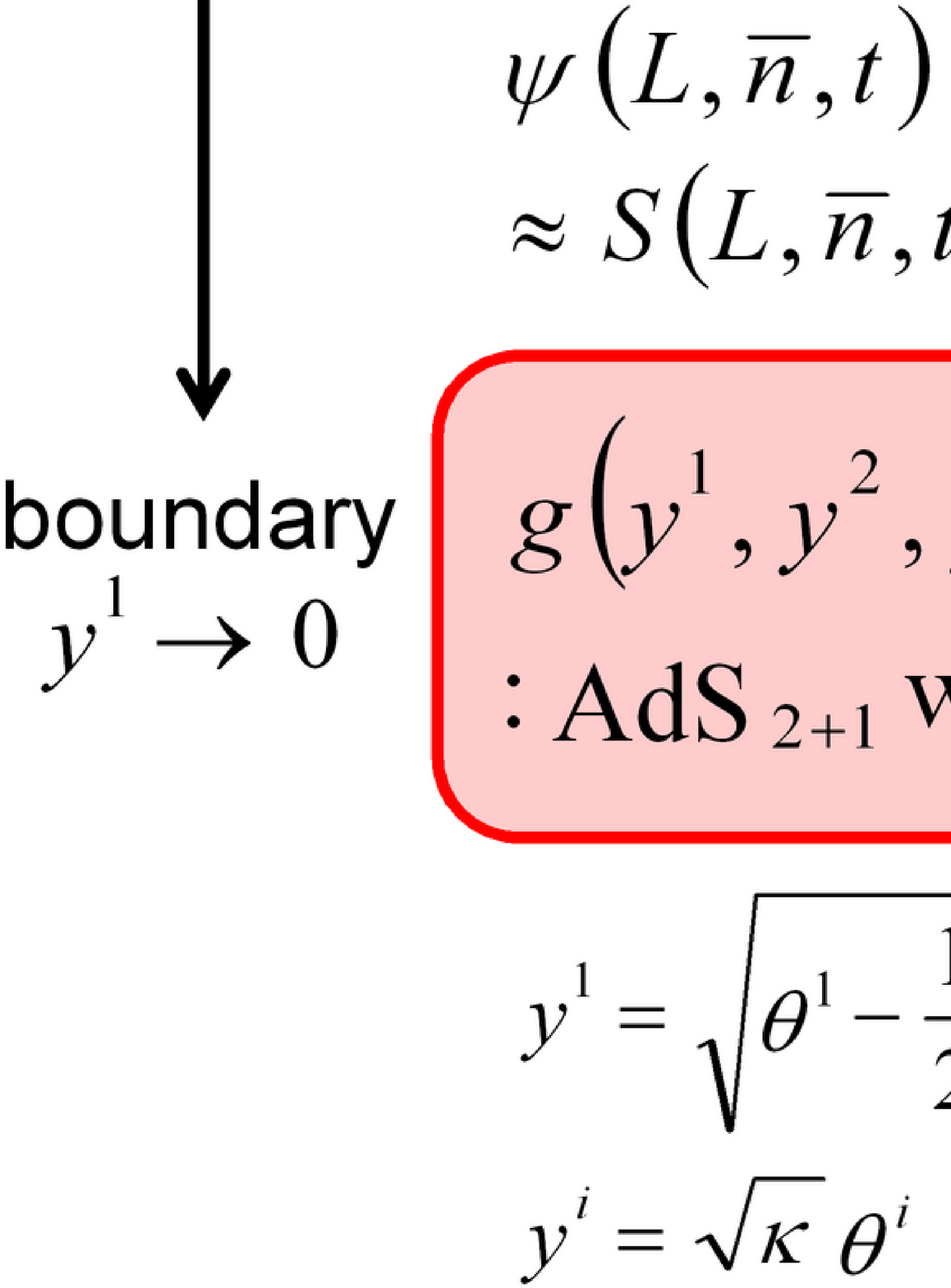}
\end{center}
\caption{(Color online) Relation map of various quantities.}
\label{fig1}
\end{figure}

We summarize important quantities in the present theory in Fig.~\ref{fig1}. The most important finding is the presence of {\it internal structure} of the AdS/CFT correspondence. To find the correspondence, once we must go to canonical parameter space, and construct a Hessian potential. Interestingly, this internal structure is a kind of Fourier space for a pseudo-fermon system. The functional form of the Hessian potential is determined so that the Fisher metric becomes AdS. In this case, we need to perform general coordinate transformation defined by Eq.~(\ref{yy}). We pull such potential back into the original representation by $(L,\bar{n},t)$. Since the Hessian potential can be identified with the entanglement entropy, we compare the potential represented by $(L,\bar{n},t)$ with various scaling formulae of the entanglement entropy. Then, we find whether there exists a certain correspondence between AdS and CFT. Our analysis provides a positive answer to this problem. Although several procedures are somehow complicated, we can also find a bulk/boundary correspondence in a sense that the original quantum system lives in the boundary of the AdS spacetime: if we go to $L\rightarrow\infty$ in the original representation of the quantum side, then we approach the boundary $y^{1}\rightarrow 0$ of the Poincare disk in the classical AdS spacetime. In usual information geometry, the concept of the bulk/boundary correspondence is not included, but the present analysis suggests the presence of such correspondence. We believe that our approach is essentially equal to the standard notion of the AdS/CFT correspondence.

Finally, we note that a recent paper on transformation of non-commutative geometry into classical geometry seems to use a method similar to the Fisher metric~\cite{Ishiki}. This similarity is also an evidence of powerfulness of information-geometrical approaches when we are going to smear out quantum mechanically fluctuating data to construct a classical spacetime.

I acknowledge Jonathan Shock for enlightening discussions during my stay in South Africa and his stay in Sendai. I also acknowledge Yoichiro Hashizume for discussion. This work was supported by JSPS KAKENHI Grant Number 15K05222.

\end{document}